# Comment on "Comments regarding "Transonic dislocation propagation in diamond" by Katagiri, et al. (Science 382, 69-72, 2023)" by Hawreliak, et al. (arXiv:2401.04213)

Kento Katagiri[1,2,3,4,5]*, Tatiana Pikuz[6], Lichao Fang[3,4,5], Bruno Albertazzi[7], Shunsuke Egashira[2], Yuichi Inubushi[8,9], Genki Kamimura[1], Ryosuke Kodama[1,2,6], Michel Koenig[1,7], Bernard Kozioziemski[10], Gooru Masaoka[1], Kohei Miyanishi[9], Hirotaka Nakamura[1], Masato Ota[2], Gabriel Rigon[11], Youichi Sakawa[2], Takayoshi Sano[2], Frank Schoofs[12], Zoe J. Smith[13], Keiichi Sueda[9], Tadashi Togashi[8,9], Tommaso Vinci[7], Yifan Wang[3,4,5], Makina Yabashi[8,9], Toshinori Yabuuchi[8,9], Leora E. Dresselhaus-Marais[3,4,5], and Norimasa Ozaki[1,2]

Affiliations:

[1]Graduate School of Engineering, Osaka University; Suita, 565-0871, Japan.

[2]Institute of Laser Engineering, Osaka University; Suita, 565-0871, Japan.

[3]Department of Materials Science & Engineering, Stanford University; Stanford, 94305, USA.

[4]SLAC National Accelerator Laboratory; Menlo Park, 94025, USA.

[5]PULSE Institute, Stanford University, Stanford, 94305, California, USA

[6]Institute for Open and Transdisciplinary Research in Initiatives, Osaka University; Suita, 565-0871, Japan.

[7]LULI, CNRS, CEA, Ecole Polytechnique, UPMC, Univ Paris 06: Sorbonne Universites, Institut Polytechnique de Paris; Palaiseau, F-91128, France.

[8]Japan Synchrotron Radiation Research Institute; Sayo, 679-5198, Japan.

[9]RIKEN SPring-8 Center; Sayo, 679-5148, Japan.

[10]Lawrence Livermore National Laboratory; Livermore, 94550, USA.

[11]Department of Physics, Nagoya University; Nagoya, 464-8602, Japan.

[12]United Kingdom Atomic Energy Authority, Culham Science Centre; Abingdon, OX14 3DB, United Kingdom.

[13]Department of Applied Physics, Stanford University; Stanford, 94305, USA.

*Corresponding author. Email: kkatagiri@ef.eie.eng.osaka-u.ac.jp

**Abstract:**

**In their comment (*1*), Hawreliak *et al.* claims that our observation of stacking fault formation and transonic dislocation propagation in diamond (*2*) is not valid as they interpret the observed features as cracks. In this response letter, we describe our rationale for interpreting the observed features as stacking faults. We also address other points raised in their comments, including the clarifications of how the results of Makarov *et al.* (*3*) are not in conflict with our study.**

**X-ray phase-contrast radiography**

Hawreliak et al. (*1*) raised several points about our *in-situ* X-ray imaging data of laser-shocked single-crystal diamond data (*2*). One of Hawreliak et al.'s points is about the naming of the experimental schematic. We called their schematic as *X-ray radiography* which usually mean the given contrast is based on the absorption, while our images capture some phase contrast effect (*2*). The reason we called the schematic as *X-ray radiography* in our paper is, as we already described in our paper, to distinguish it from the previous *phase-contrast imaging* used for shock experiments which set the distance between the sample and detector much longer than typical *radiography* experiments in order to enhance the phase-contrast signals in the images. For example, Seiboth et al. (*4*), Schropp et al. (*5*), and Brennan Brown et al. (*6*) set the distance 4.8 m and 4.2 m, and 3.9 m, respectively for their X-ray *phase contrast* imaging experiments, while the distance in our experiments was 0.112 m (*2*). In classical optical phase contrast imaging, absorption based contrast is removed from the image signal via phase masks placed in a Fourier plane; in X-ray science, phase contrast is usually introduced *in addition* to amplitude contrast via propagation effects. The contrast observed in our radiography indeed shows some phase contrast effects, as it is common to observe refraction in radiography using coherent X-ray radiation for characteristic features in the sample that are sufficiently small to make the Fresnel number near 1, as described in our paper (*2*). Additionally, Hawreliak et al. claimed that Makarov et al. (*3*) called their schematic, which is almost identical to the one used in (*2*), as *phase contrast imaging* but Makarov et al. (*3*) calls it *phase-contrast radiography* in their abstract. It is natural for Makarov et al. to shorten it to *phase-contrast* as their simulations and interpretation focus on the phase-contrast effect, not on the absorptive effects. In this response, we will call the schematic used in (*2*) *X-ray phase-contrast radiography* for clarity.

**Ductility in diamond**

Hawreliak et al. (*1*) also stated that the linear image features we observed behind the plastic shock wave front [Fig. 2 in ref (*2*)][Fig. 1B] are cracks, not stacking faults, because diamond is a brittle material that does not exhibit ductility. Kanel et al. stated in their textbook as "*One should say that no absolutely brittle or absolutely ductile materials exist*" (7), and this applies to any materials including diamond (*8-12*). Also, there is a significant body of work supporting initially brittle materials exhibiting ductility during both static and shock compression (*13-15*).

Brittle to ductile transitions have been characterized across materials systems for several decades. The effects of temperature and pressure cause the additional energy required to activate dislocation motion to decrease, generating a brittle-to-ductile transition. This effect has been demonstrated in tungsten (*16*) and steels (*17*) and has been discussed extensively in regard to silicon (*18*). In the context of ultrafast dislocations, Hahn et al. performed molecular dynamics simulation and observed transient supersonic dislocation motions in silicon, which is brittle at ambient conditions (*19*).

The claim that diamond cannot exhibit ductility has similarly been demonstrated to be invalid. The preliminary study of Weidner et al. (*8*) observed significant ductile deformation in diamond at a static pressure of 10 GPa and temperature of 1000 °C. Their findings were later supported by another group (*20*) who observed significant plastic flow (enabled by ductility) in polycrystalline diamond at a pressure of 3.5 GPa and at temperatures above 1000 °C. Brookes *et al.* also performed static loading experiments on diamonds with soft contact approach and found that the onset of plastic deformation occurred at 770°C and 12 GPa for type Ib synthetic diamond and 950°C and 9.5 GPa for type Ia natural diamond (*21*). Note that the shock-induced temperatures and pressures achieved in our study were 1150 K (877 °C) 184 ± 16 GPa for [110] shock direction and 560 K (287 °C) and 92 ± 15 GPa for [100] shock direction (*2*). In addition to the temperature and pressure, strain rate is also known to affect the brittle-ductile transition. Kalthoff observed a transition from brittle to ductile in high strength steel by raising the strain rate (*22*). Since the strain rate of laser-shock deformation is high ($\geq 10^8$ s$^{-1}$), it is reasonable to think that the diamond should show some ductility under the shock-conditions achieved in (*2*).

Hawreliak *et al.* (*1*) further claim that the laser-shock experiments on single crystal diamond performed by McWilliams *et al.* (*23*) clearly showed that shock-induced inelastic deformation in diamond is incompatible with an elastic-plastic response and requires, instead, a brittle fracture description. This is used to explain their claim that our dislocation and stacking-fault interpretation in shocked single crystal diamond has already been demonstrated as invalid. McWilliams *et al.*, however, never stated this in their paper and their data in fact shows the response of single crystal diamond shocked to stresses beyond its Hugoniot elastic limit (HEL) falls somewhere between elastic-plastic (ductile) and elastic-isotropic (brittle) response. Also, *in-situ* X-ray diffraction measurements of shocked micro-polycrystalline diamond reported by MacDonald *et al.* (*24*) further suggest that some strength remains in diamond when shocked beyond the HEL, indicating the presence of some ductility. Hawreliak *et al.* (*1*) also mentioned plate impact experiments from their group (*25*) that show a brittle response in single crystal diamond shocked to stresses beyond the HEL. The reason for the difference between their plate impact experiments (*25*) and the laser-shock experiments (*23,24*) remains unclear. Note that significant uncertainties on elastic-plastic or elastic-inelastic deformation responses of solids are recognized (*7,23,26*) even without the use of our XFEL imaging. However, if the difference in the shock-drivers and resulting time-scales are the reason, it is reasonable to assume the diamond observed in (*2*) would exhibit the deformation behaviors with ductility as seen in the laser-shock experiments of (*23,24*).

Hawreliak *et al.* (*1*) modeled a cone shaped crack in diamond and simulated a phase contrast image of that to support their claim of crack propagation in diamond (Fig. 1D). Their simulation results (Fig. 1E,F), however, is inconsistent with the X-ray images observed in (*2*) (Fig. 1B-C) and the shape of the crack cone they modeled is not realistic for following three reasons. First, Hawreliak *et al.*'s simulation for [100] probing direction observed a pair of diagonally propagating lines (Fig. 1F) which are clearly absent in our experiment (Fig. 1C). This inconsistency indicates the diagonal lines seen in our experiments (Fig. 1B) are from planer features formed along {111} planes of the diamond, not from a cone-shaped crack. Second, Hawreliak *et al.*'s simulation images do not show the cross-section of the lines observed in our experiments. In our experiments, the two-sets of lines are observed along two different directions, forming a crossed region near the bottom surface of the diamond. The crossed region is not reproduced in Hawreliak *et al.*'s simulations, as, in our understanding, their cone shaped crack model cannot have such crossed region of cracks. Also, Hawreliak *et al.*'s simulation suggests the cracks appear as short and discontinuous lines instead of the continuous lines observed in our experiments. Third, if the diagonal lines observed in our experiments are from the single cone modeled by Hawreliak *et al.*, the leading edges of the diagonal lines would form a horizontal line as seen in their simulations (Fig. 1E). This is inconsistent with our X-ray images showing the leading edge of the diagonal lines forming a curvature along the plastic wavefront (guided by yellow dashed curve in Fig. 1B). Note that formation of opening (mode-I) cracks or other types of voids in solids typically require tensile stress that is not present during shock compression. As such, the only region capable of producing cracks in our shock configuration would be the region at the edges of the cone, where edge rarefaction waves propagate in from the sides. This indicates that, together with the third point, cracks would not reach to the curved plastic wavefront where tensile stress due to the rarefaction waves is absent. Thus, the crack-cone modeled in Hawreliak *et al.* (*1*) is invalid as a descriptor for the experimental results in our study (*2*).

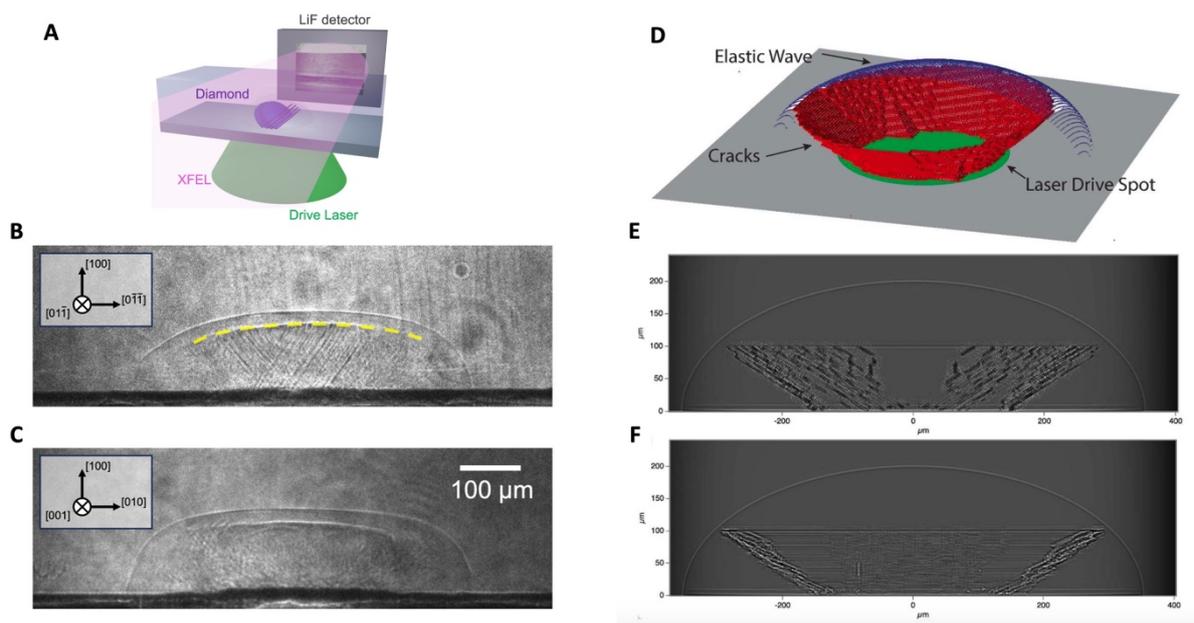

**Fig. 1** Comparison between the experimentally obtained X-ray images from shocked single-crystal diamonds in our study (*2*) (**A-C**) and Hawreliak *et al.*'s simulation results of a model assuming crack propagation in diamond (*1*) (**D-F**). The images in D-F are reproduced from ref (*1*) without any modifications, as per the terms of the Creative Commons CC BY-NC-ND 4.0 license (https://creativecommons.org/licenses/by-nc-nd/4.0/). Note that the diamond is shocked along [100] for the shown experimental data (**B**, **C**) while [110] is assumed for the simulations. Experimental data for [110] shock is also available in (*2*) but only for a limited X-ray projection direction.

Our interpretation of the diagonal lines observed in (*2*) as stacking faults has been based on eliminating other possibilities such as cracks, amorphous bands, and phase transformations known in diamond. Crack propagations would not result in the observed image as described above. Formation of amorphous bands or diffusionless phase transformation would potentially give the observed contrast (*27*), but those would not occur at the achieved pressures and temperatures because of very high phase stability of diamond (*28,29*). We do note that planar defects other than stacking faults (c.f. microtwins (*20*)) have been observed in diamond, however, the formation of these features also require that the leading edge upon propagation must arise from dislocation motion, similar to stacking faults. While other modes of shock-induced plasticity would support our conclusion of transonic dislocation motion in diamond, recent theory studies suggest the stacking-fault mechanism. A recent first-principles calculation study using density functional theory reported by Guo *et al.* showed that, the nucleation of partial dislocations with Burgers vector of 1/6 <112 > on {111} plane is energetically more favorable than the nucleation of prefect dislocations, when the hydrostatic pressure exceeds 100 GPa in diamond (*11*). This suggests that partial dislocation propagations with stacking fault formations are preferred than full dislocation propagations in diamond when the hydrostatic component of the pressure exceeds 100 GPa, finding consistency with our interpretation of stacking fault formation in the shocked single crystal diamond. Finally, the phonon radiation features we observed are predicted to be formed when cracks or dislocations are traveling transonically (*30,31*). Since the cracks are unlikely under these conditions, the observed phonon radiation features further support the presence of transonic dislocation motion in the single-crystal diamond.

It is important to note that, while the above discussion suggests dislocation motion in the shocked diamond is dominant at least for the 8 - 16 ns delay in our study (Fig. 2 of Ref (*2*)), cracks have also been observed by the X-ray imaging at later times of ≥ 60 ns (Fig. S12H-K of Ref (*2*)). The crack features in these images are evident when rarefaction waves from both the front and rear surfaces of the diamond generate tensile stresses within the crystal.

**Visualizing stacking faults**

Transmission X-ray imaging techniques are not generally used to detect dislocations or stacking faults because of limited spatial resolution and the limited refractive index contrast from these features. The spatial resolution of the X-ray phase-contrast radiography used in (*2*) is ~1 μm (*32*) which is not high enough to detect a single dislocation or stacking fault. However, plastic flow under extreme conditions would generate banded regions of bundled dislocations or stacking faults at these high dislocation densities. As such, the *bundled stacking faults* may cause localized lattice expansion that gives rise to a refractive index change and may be detectable by X-ray phase-contrast radiography.

To verify this, we performed wavefront-propagation simulations to validate the contrast required for a simple model of diamond stacking-fault bundles to produce image features via the resulting lattice expansion. Lattice defects in crystals generate strain fields around them that span much larger length scales than the defect cores. The amplitude of elastic strains around intrinsic or extrinsic stacking faults in crystals have not been thoroughly explored, however, Chou *et al*. theoretically performed such predictions for silicon (which has the diamond crystal structure). They predicted that the interplanar distance near stacking faults increases by 0.6-0.7% to alleviate the localized forces created by the stacking faults - for both intrinsic and extrinsic stacking faults (*33*). No literature has extended this to diamond, and thus we used the silicon interplanar separation predictions around stacking faults to perform wave propagation simulations to simulate the experimentally observed features.

The simulated model has air for $|x| > 500$ μm, diamond with normal density ($\rho_0 = 3.510$ g/cm$^2$) for 500 microns $\geq |x| > 160$ μm, and a periodic array of normal diamond ($\rho_0 = 3.510$ g/cm$^2$) and 0.7% expanded diamond ($\rho = 3.485$ g/cm$^2$) diamond for $|x| \leq 160$ μm. The diamond thickness along the X-ray projection direction was set 260 μm and the model uses a multi-slice approach with steps of 1 micron through the entire 260 μm thickness, computing the updated wave field at each location. The algorithm is a paraxial approximation to the wave equation, using a Fourier Transform based convolutional approach to simulate the wave propagation between planes. The pixel size was set to 0.31 μm/pixel and a 1 μm full width at half maximum optical point-spread function was applied. In this work, stacking-fault bundle periodicities of 100, 25, 10, 2.5, 1, 0.5 μm were simulated to understand how the periodicity of the density perturbations affect to the contrast at the detector. In all cases, the density change was assumed to linearly disrupt the refractive index of the diamond, and a density of 1/2 disturbances from long-range interactions of the stacking faults was assumed through the depth.

The obtained wavefront propagation simulation results with two different periodicities of 10 and 0.5 μm are shown in Fig. 2. The strongest fringe contrast is observed when the periodicity was set to 10 μm (Fig. 2A&B). The fringes for each of the periodic structures start to overlap for periods below ~10 microns. At periodicity of 2 micron or less these fringes begin to average out and the fringes within the density perturbation band becomes invisible. We note that the contrast at the interface between the density perturbation band and the surrounding normal diamonds ($|x| = 160$ μm) remains strong even when the fringes within the periodic band ($|x| \leq 160$ μm) smoothed out. The peak-to-trough ratio of the fringes observed in our experiments is most consistent with our simulation results for the 0.5 μm stacking-fault bundle periodicity. The fringes are still observed when the density difference is decreased to 0.1% at the periodicity of 10 μm (Fig. 2D).

Our simulation results suggest that stacking faults can be observed by the X-ray phase-contrast radiography if a high enough number density and associated directionality of the 2-dimensional planes are present. The periodicity of the faulted volume is also key to the contrast and width of the linear features as it generates interference fringes that are clearly evident. The periodic density perturbation required to form fringes is around a few micrometers or less, meaning many stacking faults are bundled in each of these regions, with only a few micrometers between bundles. We note that there are many possible solutions of density profiles that can form the patterns observed in the experiment. To obtain a unique solution of the density profile, we believe simultaneously obtaining images with multiple sample-to-detector distances would help as the different distances provides images with different amplitudes of the phase-contrast effect.

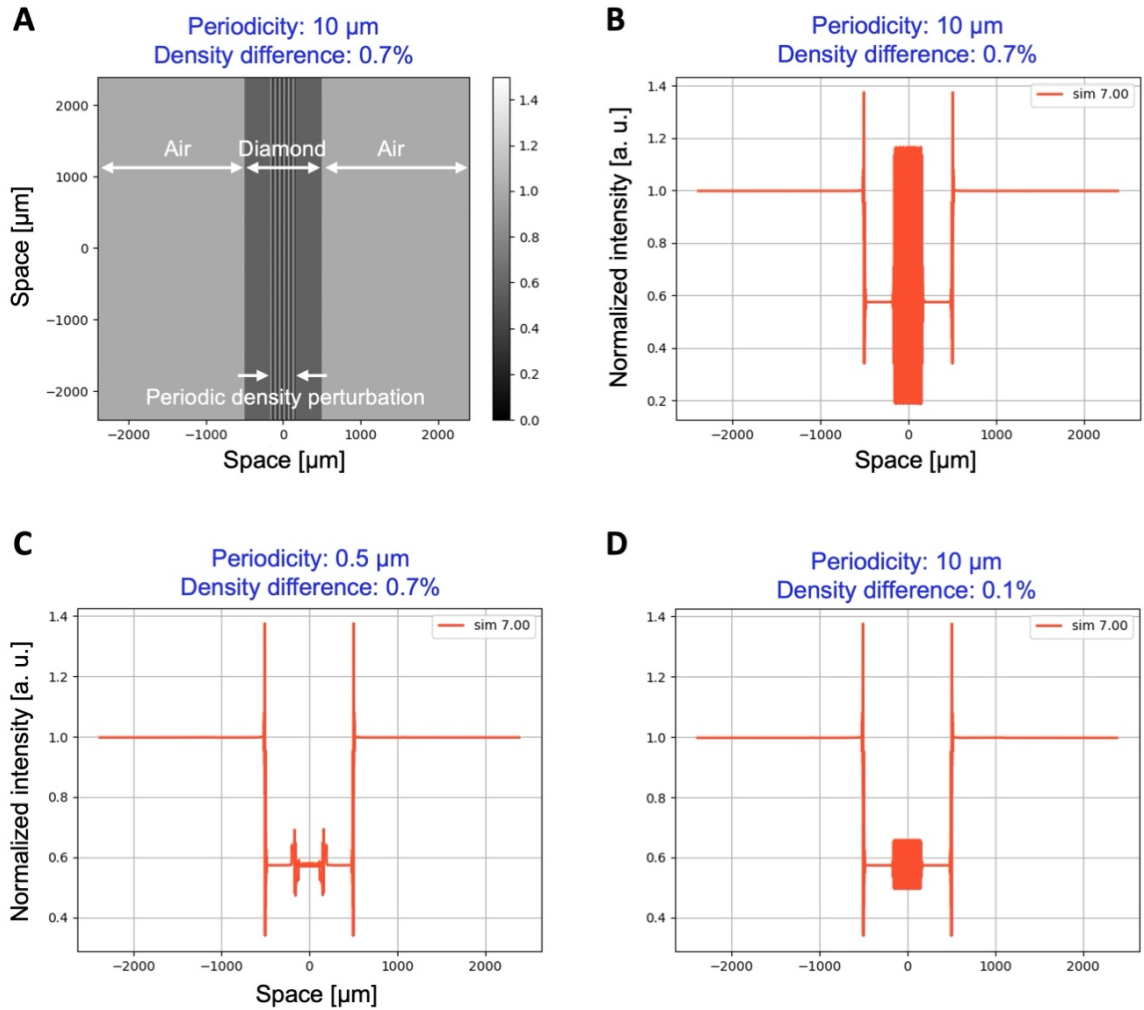

**Fig. 2** Results of wavefront propagation simulations of diamond with periodic density perturbations. Note that simulation describes wavefront propagation not of isolated stacking faults, but regions of low-density areas which could be formed by the high number density of stacking faults (see text). (**A**) 2-dimenisonal image of the simulated results with a density perturbation with periodicity of 10 μm and density difference of 0.7%. (**B**) The normalized intensity profile of A. (**C**) Results when the density perturbation periodicity is 0.5 μm and density difference of 0.7%. (**D**) Results when the density perturbation periodicity is 10 μm and density difference of 0.1%.

**Shock stress evaluation**

Another issue raised by Hawreliak *et al.* (*1*) is the shock stress evaluation method in the diamond. We estimated the shock stresses based on the shock wave velocities resulting from the measured shock wave positions at the different measured times. Then we used the existing diamond Hugoniot to convert the shock wave velocities to shock stresses. This estimation does not consider the effect of shock decaying which should be present in the shocked diamond at the time of X-ray probes (8, 12, & 16 ns) due to the 5 ns pulse duration of the drive laser. Also, the impedance mismatch of the ablator and diamond causes a back-propagating reflection wave, and when this reflection wave reaches the laser-irradiated surface of the ablator, it causes a rarefaction wave that follows the shock wave. As Hawreliak *et al.* pointed out, this stress estimation procedure is rough, and the applied laser-intensity suggests that the peak stress applied to the diamond should be higher than our estimation, meaning that some of the pressure is already released at the time of the X-ray imaging. We note, however, that even if the actual stress is higher than our estimations, the observed dislocation propagation velocities are at the predicted transonic region of diamond and thus do not affect conclusion of our paper (*2*).

**Comment on Makarov *et al*. [*Matter Radiat. Extremes* 8, 066601 (2023)]**

Hawreliak *et al.* (*1*) also pointed out that the results of our study that concludes the ductile deformation mode of single-crystal diamond during shock compression directly conflicts with the results of Makarov *et al*. (*3*) (some of the co-authors on this paper are also authors of our Science paper (*2*)) who assumed "brittle response" in shocked single-crystal diamond in their modeling. This is a misunderstanding of Makarov *et al.*'s smoothed particle hydrodynamics (SPH) simulations, and not an inconsistency between the two works.

The SPH simulations use Lagrangian particles to describe the mass, momentum, and energy conservation of the model solid continuum. The stress tensor field of the continuum is explicitly defined for calculating the momentum, angular momentum, and strain energy of the system. The materials strength in the SPH is modelled as an empirical constitutive stress-strain relation named "diamond failure model." The model is a **fitted analytical expression** (fitted with data from McWilliams *et al.*(*23*)) for calculating a degraded (softened) stress response after the Hugoniot elastic limit is reached, see the Supplementary Materials of the Makarov *et al*. (*3*). Neither crack growth nor the atomic-level failure mechanisms are considered explicitly in the model, and as mentioned above the McWilliams work does not conclude brittle fracture as the origin for the two-wave shock structure.

Although the name of the model is called "failure model," the purpose of the model is to provide a fitted analytical expression to describe the softening of the material's strength after loading beyond the elastic limit. The purpose of the Makarov *et al*. (*3*) study is to develop a simulation model able to predict the wave propagation behaviors in shocked materials. It also shows the effectiveness of the "failure model" in describing the softening of the materials strength. However, that work does not focus on the mechanism of the material's strength softening. In fact, the atomic-level mechanisms of failure model remains unclear, even for the originally developed failure model of the boron carbide ceramics (*34*). Similarly for diamond, the atomic-level mechanisms of "damaging" (softening) after elastic limit remains unclear, but many studies showed explicit evidence of mechanisms of plastic deformation as described in previous sections. As such, the related works of Makarov *et al*. are not in conflict with the results and interpretation of our study, nor with their conclusions.

**Summary**

As per the results of our optical simulations and the literature we review above, we respectfully disagree with Hawreliak *et al.* (*1*) as we find their suggestions inconsistent with our experimental observations (*2*) and a significant body of work from previous studies.